\documentclass[titlepage,12pt]{article}

\newcommand{\be}{\begin{equation}} \newcommand{\ee}{\end{equation}} 
\newcommand{\bea}{\begin{eqnarray}} \newcommand{\eea}{\end{eqnarray}}

\newcommand{\id}{\hbox{1\kern-.27em l}} 
\newcommand{\sid}{\hbox{\scriptsize1\kern-.27em l}} 
\newcommand{\pa}{\partial}  
 
\newcommand{\we}{\kern-.1em\wedge\kern-.1em} 
\newcommand{\scal}{\kern-.13em\cdot\kern-.13em}

\newcommand{\II}{I\kern-.09em I}

\begin{document} 
 
\begin{titlepage}

\rightline{\vbox{\small\hbox{YITP-SB-02-74} 
 \hbox{ITFA-2003-09}\vskip.5ex  \hbox{\tt hep-th/0302078}}} 
\vskip 1.3cm 
 
\centerline{\LARGE \bf Non-commutative superspace from string 
theory} 
 
\vskip 1.2cm  \centerline{{\Large   J. de Boer${}^a$, 
P. A. Grassi${}^b$, 
 P. van Nieuwenhuizen${}^b$} } 
\vskip 0.3cm 
\begin{center} 
{\sl ${}^a$ Institute for Theoretical Physics, University 
of Amsterdam \\[-.3ex] Valckenierstraat 65, 1018 XE Amsterdam, The 
Netherlands \\[.75ex] ${}^b$ C.N. Yang Institute for Theoretical 
Physics\\[-.3ex] State University of New York at Stony Brook, NY 
11794-3840, USA 
} 
\vskip .1in  {\small \sffamily 
jdeboer@science.uva.nl, pgrassi@insti.physics.sunysb.edu, 
vannieu@insti.physics.sunysb.edu} 
\end{center} 
\vskip 0.8cm  \centerline{\bf \large Abstract} \vskip 0.1cm 
\noindent

Turning on background fields in string theory sometimes has an 
alternative interpretation as a deformation of the target space 
geometry. A particularly well-known case is the NS-NS two form 
$B$, which gives rise to space-time non-commutativity. In this 
note we point out that this phenomenon extends to ten-dimensional 
superspace when employing a covariant quantization of the superstring,
generalizing an observation by Ooguri and Vafa in four dimensions. 
In particular, we will find that 
RR field strengths give rise to a 
non-zero $\{\theta,\theta\}$ anti-commutator, just as in four dimensions,
whereas the gravitino yields a non-zero value for $[x,\theta]$.

\end{titlepage} 
 
\section{Introduction} 
 
The idea that the coordinates of spacetime do not commute was 
proposed by Snyder in 1947 \cite{old}. When supersymmetry and 
supergravity were invented, it was a natural idea to consider also 
the possibility that the fermionic coordinates of superspace do 
not commute. In 1982 a model was constructed \cite{schw} in which 
the Dirac bracket for a fermionic point particle was proportional 
to the coordinates of spacetime 
\bea\label{tt} 
\{ \theta^\alpha, \theta^{\beta} \}_D = \gamma^{\alpha \beta}_m x^m\,. 
\eea 
Due to the composite nature of $x^m$, it was hoped that this might 
yield a granular structure of spacetime. 
 
From a more mathematical point of view, non-commuting 
(super) coordinates arise naturally in non-commutative geometry and 
in particular in rea\-lizations of quantum groups \cite{manin}. 
These realizations employ (anti)-commutators of supercoordinates 
$z^A = (x^m, \theta^\alpha)$ that are quadratic in 
supercoordinates 
\bea\label{zz} 
[ z^A, z^B \} = R^{AB}_{CD} z^C z^D\,. 
\eea 
Consistency requirements lead to a cubic equation for the matrix 
$R$, the Yang-Baxter equation. In order to be able to treat cases 
like (\ref{tt}), a linear term was added to the right hand side of 
(\ref{zz}) in \cite{bouw}, and a model was constructed in $1+1$ 
dimensional space with two $\theta$'s in which (\ref{tt}) and 
(\ref{zz}) were combined. This model still satisfied the 
Yang-Baxter equation. 
 
Non-commutative coordinates also appear in string theory. As shown 
in \cite{doug1,doug2} and elaborated upon in \cite{sw}, string 
theory in the presence of an NS-NS $B$-field admits in space-time 
an alternative formulation in terms of non-commutative geometry, 
where the coordinates satisfy 
\bea\label{xx} 
[x^m, x^n] = i\theta^{mn} \,, 
\eea 
where $ \theta^{mn} = (B^{-1})^{mn}$. Non-commutative field 
theories can be obtained by taking a suitable scaling limit where 
$\alpha'\rightarrow 0$ while scaling the space-time metric as 
$g_{ij}\sim \alpha'{}^2$ and keeping the $B$-field fixed. 
 
A natural next step is to try to obtain non-commutative fermionic 
coordinates from string theory. Indications that such a structure 
might be relevant in string theory were recently found e.g. in 
\cite{indic1,indic2}. There is a rather trivial version of 
non-commutative superspace in the literature where 
$\{\theta,\theta\}$ and $[x,\theta]$ remain zero, but $[x,x]\neq 
0$. This superspace is useful in providing a superfield formalism 
for certain supersymmetric non-commutative gauge theories, see 
e.g. \cite{zamora}, but we will be interested in the case where 
also $\{\theta,\theta\}\neq 0$ and $[x,\theta]\neq 0$. A general 
ansatz of this type was already discussed in \cite{klemm}, but our 
focus will be to obtain such deformations from string theory. 
 
Two of us already considered this problem a few years ago in 
collaboration with K.~Skenderis, in the context of the 
NSR(spinning) string. However, due to the well-known difficulties 
of implementing spacetime spinors in the NSR approach, no results 
were obtained. The Green-Schwarz superstring was also considered, 
but here the well-known problems with its covariant quantization 
precluded a sound basis to depart from. Recently, however, a 
completely covariant quantization of superstrings was developed 
\cite{pvn1,pvn2,pvn3,pvn4} which covariantizes Berkovits' pure 
spinor approach (see e.g. the review \cite{berk}). A manifestly 
super-Poincar\'e-invariant nilpotent BRST operator with a finite 
number of ghost fields was obtained, and a new definition of 
physical states was derived which yields the correct spectrum for 
the open and closed superstring at the massless and massive level. 
 
Given that there now exists a covariant quantum description of a 
string model with spacetime supercoordinates $x^m$ and 
$\theta^\alpha$, it is possible to return to the question of the 
non-commutativity of supercoordinates and base the discussion on a 
concrete consistent quantum string with manifest spacetime 
super-Poincar\'e invariance. Recently, this was done in four dimensions
by Ooguri and Vafa
\cite{oogurivafa} using the four dimensional covariant formulation
of the superstring given in \cite{bersie}\footnote{We would like to 
thank R.~Dijkgraaf for informing us of the results in \cite{oogurivafa}.}.
In particular, they found that the 
the graviphoton field strength (which sits in the RR sector of the theory)
gives rise to non-commutative fermionic coordinates, and that field theories
on such spaces are sensititve to higher order topological string
amplitudes.

Here we will generalize the calculation of $[x,x]$ and $\{\theta,\theta\}$ 
in \cite{oogurivafa} to ten dimensions. In particular,
following \cite{doug1,doug2,sw}, we 
will consider strings in arbitrary constant bosonic and fermionic 
background fields. The action is obtained by modifying the free 
action by adding the integrated $(1,1)$ vertex operator with 
constant background fields. The general form of this vertex 
operator was recently obtained in \cite{pvn2}. Then we shall 
invert the kinetic operator to obtain the propagators, and after 
imposing suitable boundary conditions, we shall derive expressions 
for the (anti)-commutators of the supercoordinates in a suitable 
scaling limit. These expressions depend on the constant background 
fields. Once we allow these background fields to become 
$x,\theta$-dependent, and expand them in terms of $x$ and 
$\theta$, we arrive at relations as in (\ref{tt}) and 
(\ref{zz}). The $\{\theta,\theta\}$ anti-commutator is given by a 
bispinor that contains all RR field strengths. In particular, 
(\ref{tt}) is obtained by taking an axion (the RR pseudoscalar in IIB 
string theory) which is proportional to $\eta_{\mu\nu} x^{\mu} 
x^{\nu}$. 
 
\section{Action} 
 
In the formulation of \cite{pvn1,pvn2,pvn3,pvn4} the fields that 
appear are the superspace coordinates $x^{\mu},\theta^{\alpha}, 
\theta^{\bar{\alpha}}$, the fields $p_{\alpha},p_{\bar{\alpha}}$ 
conjugate to $\theta$, and several ghost fields. We will restrict 
attention to the type IIB string, so that both $\theta$'s have the 
same chirality. The massless vertex operators have been described 
in \cite{pvn2} and they give rise to the linearized field 
equations of type IIB supergravity. The field equations appear as 
a set of equations for a set of superfields (see also 
\cite{berkhowe}) which are the entries of the following matrix: 
\be \label{defm} {\cal M}_{M\bar{N}} = 
\left( \begin{array}{ccc} A_{\alpha}{}_{\bar{\beta}} & 
A_{\alpha}{}_{n} & 
A_{\alpha}{}^{\bar{\beta}} \\ 
A_{m}{}_{\bar{\beta}} & A_{m}{}_{n} & 
A_{m}{}^{\bar{\beta}} \\ 
A^{\alpha}{}_{\bar{\beta}} & A^{\alpha}{}_{n} & 
A^{\alpha}{}^{\bar{\beta}} 
\end{array} \right) . 
\ee 
We will at first only be interested in vertex operators 
corresponding to constant fields and/or field strengths. To be 
precise, we will impose 
\be \label{as1} 
\partial_m A_{\alpha\bar{\beta}}= \ldots = \partial_m A^{\alpha\bar{\beta}} 
=0 . 
\ee 
In addition, we will take 
\be \label{as2} 
D_{\gamma} A^{\alpha\bar{\beta}} = D_{\bar{\gamma}} 
A^{\alpha\bar{\beta}}=0. 
\ee 
With these assumptions, the integrated vertex operator becomes 
relatively simple. 
It is given by 
\be \label{vop} 
{\cal V}_{z\bar{z}}^{(0,0)} = F_z^M F_{\bar{z}}^{\bar{N}} {\cal 
M}_{M\bar{N}} 
\ee 
where 
\be 
F_z^M = (\partial_z \theta^{\alpha}, \Pi^m_z, d_{z\alpha}) . 
\ee 
The full form of the general integrated vertex operator is quite 
complicated, but because of the assumptions (\ref{as1}) and 
(\ref{as2}) the structure simplifies quite a lot and we are only 
left with (\ref{vop}).  We see that all the terms are quadratic 
in quantum fields, including the ghosts. We have checked for 
the open string that requiring the curvature $f_{mn}$ (the lowest component of the 
superfield $F_{mn}$) and the gaugino $u^\alpha$ to be constant, and choosing 
the gauge $\theta^\alpha A_{\alpha}(x,\theta) =0$, the vertex 
operator is quadratic in $x^m, \theta^{\alpha}, p_\alpha$ and the ghost (except 
for a tadpole when the gluino is nonvanishing). We expect the same to happen for the closed 
string. This is very convenient and implies that the whole calculation 
should be identical to one done in e.g. the Berkovits formalism. 

If we use the known expansion of (\ref{defm}) in component fields 
and the equations of motion for the background fields, and add the 
integrated vertex operator (\ref{vop}) to the free world-sheet 
action, the ghost-independent part of the action becomes 
\bea 
\label{action} S = {1\over 4 \pi \alpha'} \int d^2z && \left[ \pa 
x^m \bar\pa x^n (g_{mn} - 2 \pi \alpha' b_{mn}) + p_\alpha \bar\pa 
\theta^{\alpha} + 
\pa \theta^{\bar \alpha}  \bar p_{\bar\alpha} \right. \nonumber \\ 
&& \left. + 2 \pi \alpha' 
\left( p_\alpha \psi^{\alpha}_m \bar\pa x^m + 
\pa x^m \bar\psi^{\bar\alpha}_m \bar p_{\bar\alpha} - 
p_\alpha F^{\alpha \bar\alpha} \bar p_{\bar\alpha} 
\right)\right] 
\eea 
where $\pa = {1\over 2}(\pa_\sigma + i \pa_\tau)$, 
$\bar\pa = {1\over 2}(-\pa_\sigma + i \pa_\tau)$, and 
$b_{mn} = i B_{mn}$ in Euclidean space, with $B_{mn}$ real. 
Furthermore, $p_\alpha, \bar p_{\bar \alpha}$ and 
$F^{\bar \alpha \alpha}$ are antihermitian, while 
$\theta^\alpha, \theta^{\bar \alpha}, \psi^\alpha_m$, and $\psi^{\bar\alpha}_m$ are hermitian. 
 
As a matter of convention, we put an extra factor of $2\pi 
\alpha'$ in front of the terms that came from the vertex operator. 
 
\section{Two-point functions} 
 
One way to proceed from (\ref{action}) is to integrate out 
$p_\alpha$ and $\bar p_{\bar\alpha}$. If we assume that $F^{\alpha 
\bar\alpha}$ is invertible (this condition can easily be relaxed) 
one obtains 
\bea 
S ={1\over 4 \pi \alpha'} 
\int d^2z && 
\left[ 
\pa x^m \bar\pa x^n 
\left( g_{mn} - 2 \pi \alpha' b_{mn} + 2 \pi \alpha' 
\psi^{\bar \alpha}_m F^{-1}_{\bar \alpha \alpha} \psi^\alpha_n \right) 
\right. \nonumber \\ 
&& \left. + 
\pa \theta^{\bar \alpha} F^{-1}_{\bar\alpha \alpha} \psi^{\alpha}_m \bar\pa x^m + 
\pa x^m \bar\psi^{\bar\alpha}_m F^{-1}_{\bar\alpha \alpha} \bar \pa \theta^{\alpha} + 
{1\over 2 \pi \alpha'}  \pa \theta^{\bar \alpha} F^{-1}_{\bar\alpha \alpha} \bar \pa \theta^{\alpha} 
\right] \nonumber \\ 
\eea 
 
Notice that we can redefine 
\bea 
\label{redef} \theta^{\alpha}& \rightarrow &  \theta^{\alpha} - 
2\pi\alpha' 
\psi^{\alpha}_m x^m \nonumber \\ 
\theta^{\bar{\alpha}} & \rightarrow & \theta^{\bar{\alpha}} - 
2\pi\alpha' \psi^{\bar{\alpha}}_m x^m 
\eea 
which has the effect of removing all gravitino dependence from the 
action. Without loss of generality we will therefore set the 
gravitini equal to zero and reinstate them at the end. 
 
The propagators 
\bea 
G^{MN}(z,z') = 
\left( 
\begin{array}{ccc} 
G^{mn}(z,z') &  G^{m \beta}(z,z') &  G^{m \bar\beta}(z,z') \\ 
G^{\alpha n}(z,z') &  G^{\alpha \beta}(z,z') &  G^{\alpha \bar\beta}(z,z') \\ 
G^{\bar\alpha n}(z,z') &  G^{\bar \alpha \beta}(z,z') &  G^{\bar\alpha \bar\beta}(z,z') \\ 
\end{array} 
\right) 
\eea 
are obtained by inverting the field operators $F_{MN}$ as $F_{MN} 
G^{NP}(z,z') = - \alpha' \delta_M^{~~P} \delta(z-z')$. One finds 
that (with zero gravitini) only $G^{mn},G^{\alpha 
\bar\beta},G^{\bar \alpha \beta}$ have sources, and satisfy 
\bea 
&& 
\pa \bar\pa \, G^{mn}(z,z') = - {1\over 2}\alpha' g^{mn} \delta^2(z-z') \nonumber \\ 
&& \pa \bar\pa \, G^{\bar\gamma \beta}(z,z') = 2\pi \alpha'{}^2 F^ 
{\beta \bar\gamma} \delta^2(z - z') \nonumber \\ 
&& \pa \bar\pa \, G^{\gamma \bar\beta}(z,z') = - 2\pi \alpha'{}^2 
F^{\gamma \bar\beta} \delta^2(z - z')  \label{11} . 
\eea 
%
%
%
%
 
The boundary conditions which follow from the Euler-Lagrange field 
equations for $\theta^\alpha$ and $x^n$ read 
\bea\label{16} 
&&\left. \left(\pa \theta^{\bar\alpha}  -\bar \pa \theta^\alpha \right)\right|_{z =\bar z}= 0 \nonumber \\ 
&&\left. (g_{mn} - 2 \pi \alpha' b_{mn}) \bar \pa x^n - 
(g_{mn} + 2 \pi \alpha' b_{mn}) \pa x^n  \right|_{z =\bar z}= 0 \,. 
\eea 
To obtain the first condition, we assumed that $\theta^\alpha = \theta^{\bar\alpha}$ on the 
boundary; in flat space this follows from the requirement that the action be supersymmetric, and we assume 
that it continues to hold in the presence of costant background fields. 
The last boundary condition is the usual boundary 
condition \cite{sw} for the bosonic sector. The equation for 
$G^{mn}(z,z')$ in (\ref{11}) is also the standard one, and 
therefore the propagator $G^{mn}(z,z')$ has exactly the form given 
in eqns (2.3)-(2.5) in \cite{sw}.

We shall now determine $G^{\gamma \beta }(z,z')$. From (\ref{11}) 
we find 
\bea\label{g1} 
G^{\gamma \beta}(z,z') = P^{\gamma \beta} \ln(z - \bar z') + Q^{\gamma \beta} \ln(\bar z - z') \,. 
\eea 
To determine $P^{\gamma \beta}$ and $Q^{\gamma \beta}$ we use the boundary conditions 
in (\ref{16}) to deduce the following relations 
\bea\label{g2} 
\pa G^{\bar \beta \gamma} = \bar \pa G^{\beta \gamma}\,, ~~~~~~~ 
\pa G^{\bar \beta \bar \gamma} = \bar \pa G^{\beta \bar \gamma}\,. 
\eea 
Since $G^{\bar \beta \gamma}(z,z')$ is given by $\alpha' 
(2\pi\alpha') F^{\gamma \bar \beta} \ln| z-z'| + P^{\bar \beta 
\gamma} \ln(z-\bar z') + Q^{\bar \beta \gamma} \ln(\bar z - z')$, 
one finds from the first relation in (\ref{g2}) 
\bea\label{g3} 
\left. \alpha' { 2\pi\alpha' F^{\gamma \bar \beta} \over z - z'} + 
{P^{\bar 
\beta \gamma} \over z - \bar z'} - {Q^{\gamma \beta} \over \bar z 
- z'} \right|_{z =\bar z} = 0\,. 
\eea 
Hence $P^{\bar \beta \gamma} = 0$ while $Q^{\gamma \beta} = 
\alpha' (2\pi\alpha') F^{\gamma \bar \beta}$. In a similar manner 
one finds from the second relation in (\ref{g2}), using $G^{\bar 
\beta \bar \gamma} = P^{\bar \beta \bar \gamma} \ln(z-\bar z') + 
Q^{\bar 
\beta \bar \gamma} \ln(\bar z - z')$, and $G^{\beta \bar 
\gamma}(z,z') = - G^{\bar \gamma \beta}(z',z)$, 
\bea\label{g4} 
\left. {P^{\bar \beta \bar \gamma} \over z - \bar z'} - \alpha' 
{(2\pi\alpha')F^{\beta  \bar\gamma} \over \bar z' - \bar z} - 
{P^{\bar 
\beta \gamma} \over z' - \bar z'} \right|_{z =\bar z} = 0\,. 
\eea 
Hence $P^{\bar \beta \gamma} =0$ while $P^{\bar \beta \bar \gamma} = - \alpha' 
(2\pi \alpha') F^{\beta \bar \gamma}
$. Since $P^{\bar \beta \bar \gamma} = P^{\beta \gamma}$, 
we find 
\bea\label{g5} 
G^{\gamma \beta}(z,z') = - \alpha'(2\pi\alpha')F^{\beta \bar 
\gamma } \ln\left( {z -\bar z' \over \bar z - z'} \right)\,, 
\eea 
As in SW this implies that for $z$ and $z'$ approaching the boundary one has 
\bea\label{g6} 
G^{\gamma \beta}(z,z') = \pi i\alpha' (2\pi\alpha')F^{\beta \bar 
\gamma} {1\over 2} \epsilon(\tau - \tau') + {\rm regular~terms} 
\,. 
\eea 
Finally, this leads to 
\bea\label{g7} 
\{ \theta^\alpha, \theta^{\bar\beta} \} =i (2\pi\alpha')^2 
F^{\alpha \bar \beta}\,. 
\eea 
Notice that we did not have to take any particular scaling limit 
to obtain this result. Indeed, the quadratic term in $\theta$ in 
the action can be interpreted as a pure $B$-term, due to the 
anti-commutativity of $\theta$. 
 
\section{Non-commutative superspace} 
 
The final result that we have obtained reads 
\bea \label{fin1} 
[x^{m},x^{n}]  & = &  2 \pi i \alpha' \left( 
\frac{1}{g+2\pi\alpha' B} \right)^{mn}_A \nonumber \\ 
\{ \theta^\alpha, \theta^{\bar\beta} \} & = & i (2\pi\alpha')^2 
F^{\alpha \bar \beta}\, , 
\eea 
where the subscript $A$ denotes the restriction to the 
antisymmetric part of a matrix. Of course, on the boundary we have 
to impose that $\theta^{\alpha}=\theta^{\bar{\alpha}}$ so that it 
only the symmetric part of $F^{\alpha\bar\beta}$ that really 
appears in (\ref{fin1}). 
 
Next, we reintroduce the gravitino. We do this by shifting 
$\theta$ as explained above. After this shift, we no longer have 
the boundary condition $\theta^{\alpha}=\theta^{\bar{\alpha}}$ on 
the boundary, but instead we will have 
\be \label{bc} 
\theta^{\alpha} + 2\pi \alpha' \psi^{\alpha}_m x^m = \theta^{\bar 
\alpha} + 2\pi \alpha' \psi^{\bar \alpha}_m x^m 
\ee 
at the boundary. With nonzero gravitino we find 
\bea \label{fin2} 
[x^{m},x^{n}]  & = &  2 \pi i \alpha' \left( 
\frac{1}{g+2\pi\alpha' B} \right)^{mn}_A \nonumber \\ {} 
[x^m,\theta^{\alpha} ] & = & - i (2\pi\alpha')^2 \left( 
\frac{1}{g+2\pi\alpha' B} \right)^{mn}_A \psi^{\alpha}_n \nonumber 
\\ {} [x^m,\theta^{\alpha} ]  & = & - i (2\pi\alpha')^2 
\left( \frac{1}{g+2\pi\alpha' B} \right)^{mn}_A \psi^{\bar \alpha}_n \\ 
\{ \theta^\alpha, \theta^{\bar\beta} \} & = & i (2\pi\alpha')^2 
F^{\alpha \bar \beta} + i (2\pi\alpha')^3 \psi^{\alpha}_m \left( 
\frac{1}{g+2\pi\alpha' B} \right)^{mn}_A \psi^{\bar \alpha}_n \, . 
\eea 
A good scaling limit of this system is (assuming $B$ has maximal 
rank) to take $\alpha'\rightarrow 0$, to scale $g_{mn}$ as $(\alpha')^2$, 
and to keep fixed 
\be 
\theta^{mn}= (B^{-1})^{mn},\quad \Psi^{\alpha}_m \equiv 2\pi 
\alpha' \psi^{\alpha}_m , \quad \Psi^{\bar\alpha}_m \equiv 2\pi 
\alpha' \psi^{\bar\alpha}_m , \quad {\cal F}^{\alpha \bar \beta} 
\equiv (2\pi\alpha')^2 F^{\alpha\bar\beta}. 
\ee 
Then the (anti)commutators reduce to 
\bea 
\label{fin3} [x^{m},x^{n}]  & = &    i \theta^{mn} \nonumber \\ {} 
[x^m,\theta^{\alpha} ] & = & - i \theta^{mn} \Psi^{\alpha}_n 
\nonumber 
\\ {} [x^m,\theta^{\alpha} ]  & = & - i \theta^{mn} \Psi^{\bar \alpha}_n \\ 
\{ \theta^\alpha, \theta^{\bar\beta} \} & = & i {\cal F}^{\alpha 
\bar 
\beta} + i  \Psi^{\alpha}_m \theta^{mn} 
\Psi^{\bar \alpha}_n \, . 
\eea 
 
This is our final result for the non-commutative superspace as 
obtained from string theory. 
 
\section{Non-constant background fields} 
 
So far we have taken the background fields to be constant. One can 
in principle also consider non-constant background fields. For 
commuting coordinates, this was studied in detail by Kontsevich 
\cite{kont}. If we put $[x^{\mu},x^{\nu}]=i\theta^{\mu\nu}(x)$, 
the Jacobi identities impose certain constraints on $\theta^{mn}$. 
If these Jacobi identities are satisfied, the commutators can be 
used to define a generalization of the Moyal $\ast$-product. 
 
The relation between the work of Kontsevich and string theory was 
clarified in \cite{cf}. From the bosonic string one can extract a 
topological theory that upon quantization produces the generalized 
$\ast$ product. This generalized product is therefore present in 
string theory, but emerges most clearly in a topological limit. If 
we do not take the topological limit, the full structure of the 
theory is more complicated. 
 
It would be interesting to work out corresponding statements for 
non-commutative superspaces with non-constant (anti)-commutators. 
Again, Jacobi identities will constrain the space-time fields. In 
addition, one has to worry whether or not the space-time fields 
solve the string theory equations of motion 
(for some discussion, see \cite{dp}). Nevertheless, we 
expect that also in this case, non-commutative superspace with 
space-time dependent (anti)-commutators describes a sector of 
string theory in the corresponding background fields. We leave a 
detailed study of these issues to future research. 
 
\section{Conclusions} 
 
In conclusion, we have seen that non-commutative superspace 
appears quite naturally in string theory. There are many 
interesting problems and generalizations that one may consider. 
 
The world-volume theory for D-branes is described, in superspace, 
by a $\kappa$-symmetric world-volume theory. Perhaps these can be 
generalized to non-commutative superspaces. If so, there may also 
exist a generalization of the Seiberg-Witten map \cite{sw} that 
related the commutative action with explicit background fields to 
the non-commutative action with no superfields. 
 
For special choices of (space-time dependent) backgrounds, 
particularly interesting non-commutative superspaces may appear. 
For example, as discussed in the introduction, one may try to 
obtain a structure of the form 
\be \label{c1} 
\{ \theta^{\alpha},\theta^{\beta} \} = x^{\mu} 
\gamma_{\mu}^{\alpha\beta} 
\ee 
which is similar to the bracket obtained in \cite{schw}. In view 
of our results, brackets of the form (\ref{c1}) can appear by 
considering e.g. an axion configuration (which is the RR scalar) 
of the form $a\sim \eta_{\mu\nu} x^{\mu} x^{\nu}$. This is not an 
axion background that to our knowledge has any special meaning, 
and in fact appears rather problematic. Still, it would be 
interesting to explore this further. 
 
On a more formal level, one would like to understand better the 
geometrical structure of non-commutative superspace, and which 
physical theories allow an extension to this space. 
 
It is also important to realize that in the presence of background 
fields, supersymmetry will generally be broken. In e.g. type IIB 
string theory on a Calabi-Yau, the supersymmetry breaking appears 
through the superpotential $\int G\wedge \Omega$, with G the RR 
three-form field strength, and $\Omega$ the holomorphic three-form 
\cite{rr1,rr2,rr3}. We can alternatively study whether 
supersymmetry is broken directly in non-commutative superspace, by 
looking at whether the deformation is compatible with global 
supersymmetry. It would be interesting to understand more directly 
the relation between these two ways of breaking supersymmetry. 
Ultimately this may lead to a useful novel mechanism of 
supersymmetry breaking.

\section{Acknowledgement} 
 
We would like to thank Robbert Dijkgraaf for discussions and for
communicating some of the results in \cite{oogurivafa}.
In addition, we would like to thank G. Policastro for working out the vertex 
for constant backgrounds. The research of 
JdB is partially supported by the stichting FOM. Research of PAG and 
PvN is partially supported by NSF Grant PHY-0098527.

\end{document}